\documentclass[runningheads]{llncs}
\usepackage{graphicx}
\usepackage{times}
\usepackage{epsfig}
\usepackage{graphicx}
\usepackage{amsmath}
\usepackage{amssymb}
\usepackage{multirow}
\usepackage{multicol}
\usepackage{tabularx}
\usepackage{booktabs}
\usepackage{mathtools}
\begin{document}
\title{Improving Interpretability of Deep Neural Networks in Medical Diagnosis by Investigating the Individual Units}
%
%
\author{Woo-Jeoung Nam\inst{1} \and
Seong-Whan Lee\inst{2}}
\authorrunning{W-J. Nam et al.}
%
\institute{Department of Computer and Radio Communications Engineering\\ \and
Department of Artificial Intelligence\\Korea University, Seoul, Republic of Korea\\
\email{\{nwj0612,sw.lee\}@korea.ac.kr}}
\maketitle              
\begin{abstract}
As interpretability has been pointed out as the obstacle to the adoption of Deep Neural Networks (DNNs), there is an increasing interest in solving a transparency issue to guarantee the impressive performance. In this paper, we demonstrate the efficiency of recent attribution techniques to explain the diagnostic decision by visualizing the significant factors in the input image. By utilizing the characteristics of objectness that DNNs have learned, fully decomposing the network prediction visualizes clear localization of target lesion. To verify our work, we conduct our experiments on Chest X-ray diagnosis with publicly accessible datasets. As an intuitive assessment metric for explanations, we report the performance of intersection of Union between visual explanation and bounding box of lesions. Experiment results show that recently proposed attribution methods visualize the more accurate localization for the diagnostic decision compared to the traditionally used CAM. Furthermore, we analyze the inconsistency of intentions between humans and DNNs, which is easily obscured by high performance. By visualizing the relevant factors, it is possible to confirm that the criterion for decision is in line with the learning strategy. Our analysis of unmasking machine intelligence represents the necessity of explainability in the medical diagnostic decision.

\end{abstract}
Deep Neural Networks (DNNs) play an important role in improving empirical performance in various computer vision tasks such as image classification \cite{krizhevsky2012imagenet,he2016deep}, object detection \cite{szegedy2013deep}, human action recognition \cite{yang2007reconstruction,roh2007accurate,ji20133d,ahmad2006human}, and medical diagnosis \cite{cheng2016computer,liu2014early}. However, the lack of interpretability prevents many DNN models from being applied mission-critical systems including medical diagnosis, military, and finance. Despite the remarkable successes of Computer-aided detection (CADe) and Computer-aided diagnosis (CADx) \cite{litjens2017survey,kooi2017large,bulthoff2003biologically}, the adoption in the real world is still hindered by the ambiguity in understanding the diagnostic decision. To overcome this limitation, many researches have been studied to address the lack of transparency in DNNs.

In explaining the decision of DNNs as a way of decomposition, the contributions of individual neurons are propagated backward through the weights, resulting in a redistribution of relevance in the pixel space. Sensitivity analysis \cite{baehrens2010explain} implies the factors which reduce/increase the evidence for the predicted results. Layerwise relevance propagation (LRP) \cite{bach2015pixel} is a method for decomposing the output prediction by fully redistributing the relevance throughout the layers. Deep Taylor Decomposition \cite{montavon2017explaining} is a theoretical extension of LRP to interpret the decision by utilizing the Taylor expansion and root point concept. DeepLIFT \cite{shrikumar2017learning} propagates the differences of contribution scores between the activated neurons and their reference activation. Recently proposed Relative attributing propagation (RAP) \cite{nam2019relative} is a method for decomposing the positive (relevant) and negative (irrelevant) relevance to each neuron, according to its relative influence among the neurons. By changing the perspective from value to influence, it shows a clear distinction of relevant/irrelevant attributions with a high objectness score. Relative sectional propagation (RSP) \cite{nam2021interpreting} aims to decompose the output predictions of DNNs with class-discriminativenss by setting the hostile activations of neurons with respect to the target class.

In the visualization of disease aids radiologists, CAM \cite{zhou2016learning}, generating class activation mappings to indicate the discriminative regions, is widely utilized in the medical domain to localize diversity thoracic disease. Despite the advantages of simple implementation and class discriminations, a broad area of visualization that occurred by the expansion of compressed feature maps is pointed out as the main obstacle to explain the predictions precisely. To localize the lesion areas more clearly, image-to-image translation methods based on generative model have been studied in the medical field \cite{seah2019chest,siddiquee2019learning,taghanaki2019infomask,zhang2019biomarker} by visualizing the different factors of inter-domains while maintaining the original subject. VA-GAN~\cite{baumgartner2018visual} and Fixed-point GAN~\cite{siddiquee2019learning} synthesize alzheimer's disease images using 3d brain magnetic resonance imaging (MRI). However, there are limitations that GAN network itself is not completely explainable, and can not guarantee generality in fields without sufficient localization ground truths of lesion areas. 

To resolve this problem, we address the efficiency of recent attribution techniques to interpret the diagnostic decisions. These algorithmic approaches investigate the trained network itself without any additional supervision. Fig.\ref{fig1} illustrates the intuitive examples of visual explanations in this work. The main contributions of this work are as follows:

\begin{figure}[!t]
\includegraphics[width=\textwidth]{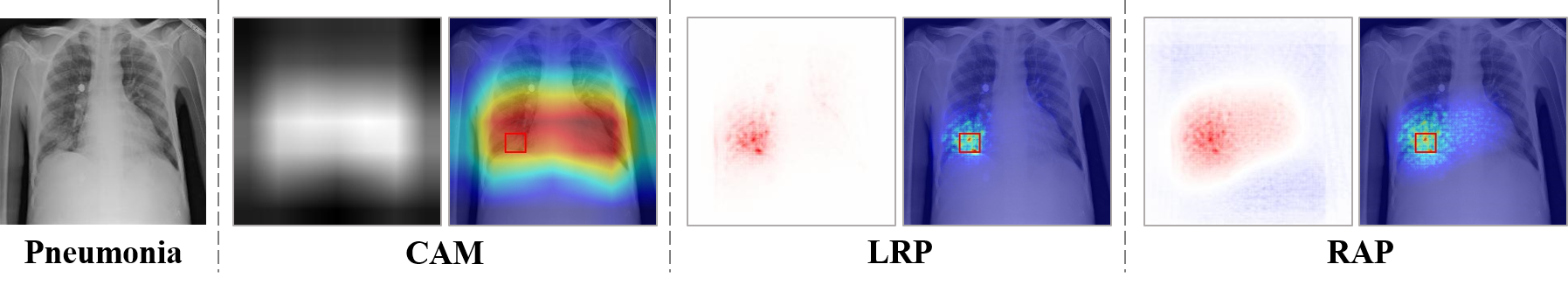}
\caption{The intuitive comparison of explaining methods: CAM, LRP, and RAP. For each method, the left and right images show the relevance heatmap and visual explanation, respectively. Red: high relevant, Blue: low relevant.} \label{fig1}
\end{figure}
\begin{itemize}
\item We demonstrate the efficient methods to interpret the diagnostic decision of DNNs by utilizing visual explaining techniques: LRP and RAP, which are applicable to any fully trained networks. Without any supervision of the area of lesions, decomposing the output predictions of classification networks makes it possible to clearly localize and illustrate the crucial factors for their diagnosis. Our experiments using chest X-ray datasets show that the recent attribution techniques provide more sufficient guarantee of performance compared to the existing widely used CAM.
\item 
We demonstrate the experiment of the inconsistency between human intention and DNNs by training binary classification tasks: Normal or Pneumonia. We utilize general techniques in machine learning fields, and the networks have proper performance. However, the visual explanation shows the misalignment between the direction of the learning strategy and the actual criteria of the classification. We analyze the phenomenon of inconsistency and add a voice to the necessity of interpretability. 

\end{itemize}

\begin{figure}[!t]
\includegraphics[width=\textwidth]{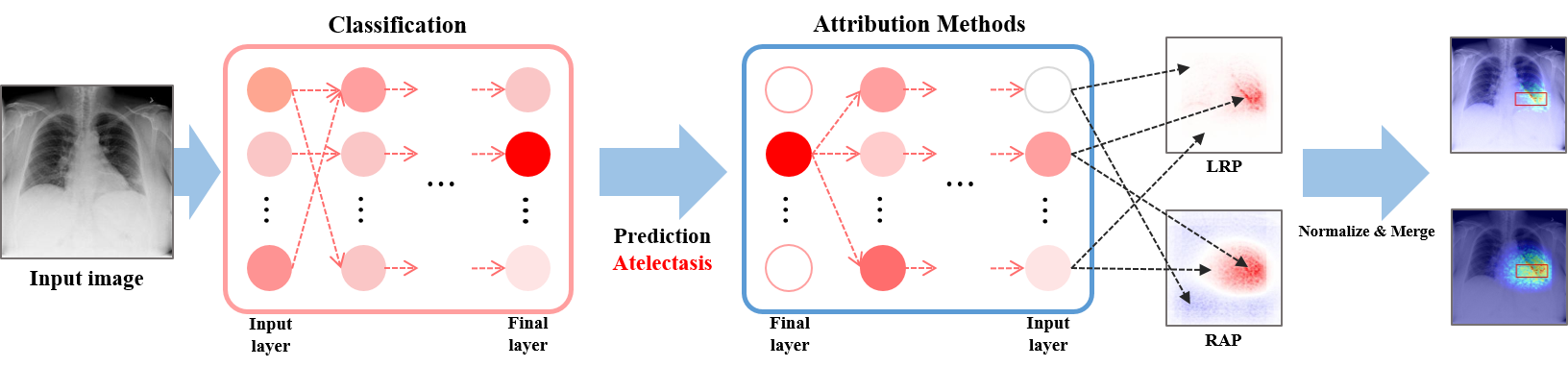}
\caption{The overview of the visual explanation procedure. Attribution methods: LRP and RAP are the decomposing procedure in a backward pass after the network is fully trained. Relevance, corresponding to output prediction, is propagated through trained weights and activated neurons.} \label{fig2}
\end{figure}

\section{Attribution Methods}
In this section, we introduce notations and attribution methods: LRP and RAP, which are closely related to each other, but have different perspective and algorithm. The overview of decomposition and visualization is illustrated in Fig.\ref{fig2}. For input \(x\), we denote the letter \(f(x)\) the value of the network output before passing through the classification layer, such as sigmoid and softmax layer. \(R\) represents the input relevance for the attributing procedure which is same as the value of \(f(x)\) of the prediction node. \(w_{ij}^{(l,l+1)}\), \(b_j^{(l,l+1)}\) and \(a(\cdot)\) denote the weight, bias and activation function between layer \(l, l+1\), respectively. \(m_{j}^{(l+1)}\) is the value of neuron after applying activation function. The signs of positive and negative values are denoted by \(+\) and \(-\).

\subsubsection{Layerwise Relevance Propagation}
 The principle of LRP \cite{bach2015pixel} is to find the parts with high relevance in the input by propagating the result from back (output) to front (input). The algorithm is based on the conservation principle, which maintains the relevance in all layers: from input to output. 
\begin{equation}
    \sum_i R_{i}^{(l)} = \sum_j R_{j}^{(l+1)}
    \label{eq:1}
\end{equation}
In various LRP versions introduced in \cite{bach2015pixel}, we utilize LRP-\(\alpha\beta\), which separates the positive and negative activations during the relevance propagation process while maintaining the conservation rule (\ref{eq:1}).
\begin{equation}
    R_{i}^{(l)} = \sum_{j} \left(\alpha\cdot\frac{z_{ij}^+}{\sum_{i}z_{ij}^+}-\beta\cdot\frac{z_{ij}^-}{\sum_{i}z_{ij}^-}\right)R_{j}^{(l+1)}
    \label{eq:2}
\end{equation}
In this rule, \(z_{ij}^+ + z_{ij}^- = z_{ij}\) and \(\alpha - \beta = 1\). The propagated attributions are allocated to the pixels of the input image while indicating how relevant to output prediction. In this paper, the function parameters are set as \(\alpha=1, \beta=0\).
\subsubsection{Relative Attributing Propagation}
RAP \cite{nam2019relative} decomposes the output predictions of DNNs in terms of relative influence among the neurons, resulting in assigning the relevant and irrelevant attributions with a bi-polar importance. By changing the perspective from value to influence, the generated visual explanations show the characteristics of strong objectness and clear distinction of relevant and irrelevant attributions. The algorithm has three main steps: (i) absolute influence normalization, (ii) deciding the criterion of relevance and propagating in a backward pass, and (iii) uniform shifting for changing the irrelevant neurons to negative.

Absolute Influence Normalization is the process applied in only first backward propagation for changing the perspective to the neuron from the value to the influence. From the output prediction node \(j\) in layer \(q\), the relevance is allocated into the penultimate layer \(p\) according to their actual contribution in a forward pass.
\begin{equation}
\begin{gathered}
    R_{i}^{(p)} = \left(\sum_{i}z_{ij}^++\sum_{i}z_{ij}^-\right) * R_{j}^{(q)}
\end{gathered}
\label{eq:3}
\end{equation}
To approach as an influence perspective, positive or negative relevance values allocated in penultimate layer are normalized by the ratio of the absolute positive and negative values \(|R_{i}^{(p)+}|:|R_{i}^{(p)-}|\).
\begin{equation}
R_{i}^{\prime(p)} = |R_{i}^{(p)}| * \frac{\sum_{i}R_{i}^{(p)}}{\sum_{i}|R_{i}^{(p)}|}
\label{eq:4}
\end{equation}
This process makes the neurons to be allocated as relative importance to the output prediction, from highly influenced to rarely influenced. For the next steps, i.e., the attributing procedure from the penultimate layer to the input layer, eq.(\ref{eq:5}) and (\ref{eq:6}) are repeated in each layer with changing low influential neurons to negative relevance.
\begin{equation}
\begin{gathered}
\bar R_{i}^{(l)} = 
\sum_{j}
\Bigl(\frac{z_{ij}^+}{\sum_{i}\left(z_{ij}^+\right)} R_{j}^{(l+1)} + \frac{z_{ij}^{-}}{\sum_{i}\left(z_{ij}^{-}\right)}\left(R_{j}^{(l+1)} * \frac{\sum_{i}|z_{ij}^-|}{\sum_{i}\left(|z_{ij}^+| + |z_{ij}^-|\right)}\right)\Bigr)\\
\end{gathered}
\label{eq:5}
\end{equation}
\begin{equation}
\Psi_{i}^{l} = 
\begin{cases}
\sum_{i} \left(\bar R_{i \in \mathcal{N}}^{(l)}\right) * \frac{1}{\Gamma}  &,    \text{ $m_{i}^{(l)}$ is activated} \\
0 &,    \text{ otherwise} 
\end{cases}
,\quad R_{i}^{(l)} = \bar R_{i}^{(l)}-\Psi_{i}^{l}
\label{eq:6}
\end{equation}
Here, $\Gamma$ is the number of activated neurons in each layer, and $\bar R_{i \in \mathcal{N}}^{(l)}$ denotes the relevance propagated through the negative weights, i.e. the latter parts of eq.(\ref{eq:5}). This procedure makes it possible to assign relatively irrelevant units as negative while emphasizing the important factors as highly positive. RAP also preserves the conservation rule (\ref{eq:1}).
\section{Experimental Evaluation}
\subsection{Data}
\subsubsection{NIH ChestX-ray14} NIH ChestX-ray14 Dataset \cite{wang2017chestx} comprises 112,120 X-ray images from 30,805 patients with corresponding 14 disease labels: Atelectasis, Consolidation, Infiltration, Pneumothorax, Edema, Emphysema, Fibrosis, Effusion, Pneumonia, pleural thickening, Cardiomegaly, Nodule, Mass, and Hernia. We utilize CheXNet \cite{rajpurkar2017chexnet}, which is based on DenseNet \cite{huang2017densely} and widely used in radiologist-level chest x-ray. This trained model is online available with verified performance. The average Area Under Receiver Operating Characteristic (AUROC) of this model for 14 classes is 0.843 and the AUROC of 8 classes which annotated with bounding box labels: Athelectasis, Cardiomegaly, Effusion, Infiltration, Mass Nodule, Pneumonia, Pneumothorax is in Tab.\ref{tab:1}. We utilize 984 images, annotated with bounding box, to evaluate the visual explanations for the target lesions.
\subsubsection{RSNA Pneumonia Detection}The RSNA Pneumonia Detection Challenge dataset \cite{RSNA} is a subset of 30,000 images from the NIH ChestX-ray14 dataset and labeled with two classes: Normal and pneumonia. The original purpose of this challenge is detecting the pneumonia lesions. We exclude the test data (which does not have label) and separate the training dataset into train and validation in a ratio of 9:1. We trained VGG-16 \cite{simonyan2014very}, ResNet-50 \cite{he2016deep} and DenseNet-121 networks, which are successfully settled in machine learning field with impressive performance. While this dataset is designed for localizing the lesions, we train the classification network, which is a much easier level than training the detection network. The purpose of interpreting these models is to analyze that their criterion of classification is fair compared to the human intentions. Therefore, we also compare with the assessment of CheXNet in the same experimental status to verify the analysis. The detail discussion is in section 4.

\begin{table*}[t!]
\caption{The performance of AUROC for each model we used in our experiment.}
\centering
\resizebox{\linewidth}{!}{
\begin{tabular}{c|c|c|c|c|c|c|c|c}
\toprule[0.3ex]
\textbf{NIH ChestX-ray14} & \textbf{Athelect.} & \textbf{Cardiom.} & \textbf{Effusio.} & \textbf{Infiltr.} & \textbf{Mass} & \textbf{Nodule} & \textbf{Pneunia.} & \textbf{Pneurax.}\\\midrule[0.2ex]
\begin{tabular}[c]{@{}c@{}}CheXNet \end{tabular} &
\begin{tabular}[c]{@{}c@{}}    0.829\end{tabular} &
\begin{tabular}[c]{@{}c@{}}    0.916\end{tabular} &
\begin{tabular}[c]{@{}c@{}}    0.887\end{tabular} &
\begin{tabular}[c]{@{}c@{}}    0.714\end{tabular} &
\begin{tabular}[c]{@{}c@{}}    0.859\end{tabular} &
\begin{tabular}[c]{@{}c@{}}    0.787\end{tabular} &
\begin{tabular}[c]{@{}c@{}}    0.774\end{tabular} &
\begin{tabular}[c]{@{}c@{}}    0.872\end{tabular}
\\\midrule[0.2ex]

\textbf{RSNA Pneumonia} & \multicolumn{2}{c|}{\textbf{DenseNet-121}} & \multicolumn{2}{c|}{\textbf{ResNet-50}}& \multicolumn{2}{c|}{\textbf{VGG-16}}&\multicolumn{2}{c}{\textbf{CheXNet}}\\\midrule[0.2ex]

\begin{tabular}[c]{@{}c@{}}AUROC \end{tabular} &
\multicolumn{2}{c|}{0.858} &
\multicolumn{2}{c|}{0.845} &
\multicolumn{2}{c|}{0.842} &
\multicolumn{2}{c}{0.827}
\\
\bottomrule[0.3ex]
\end{tabular}

\label{tab:1}
}
\end{table*}

\subsubsection{Assessment of Explanation}
It is difficult to judge the criterion of a better explanation because each method is designed for slightly different objectives and evaluating the quality of visualization does not have one commonly accepted measure. In analyzing radiologist-level chest x-ray images, the interpretation of diagnostic could be altered to the localization of lesions, which is the crucial evidence for deciding the patient status. Intersection of Union (IOU) is widely used in semantic segmentation or object detection tasks as an evaluation metric by computing the localization scores. The evaluation would be more accurate in case the dataset is annotated with segmentation mask, but there is a limit to annotating it in the medical domain in practice. Therefore, we utilize IOU to evaluate whether positive attributions are correctly distributed in the area of lesions (bounding box). The result we report is the localization performance of lesions without any supervision of the bounding box during the training procedure.

\begin{table*}[t]
\caption{The result of mean Intersection of Union of each method, between the bounding box and heatmaps. The threshold denotes the criterion for ignoring low relevance: $\{0\sim T\}$. The performance is the localization result without any supervision of bounding box.}
\centering
\resizebox{\linewidth}{!}{
\begin{tabular}{c|c|c|c|c|c|c|c|c|c}
\toprule[0.3ex]
\textbf{T(IOU)} & \textbf{Method} & \textbf{Athelect.} & \textbf{Cardiom.} & \textbf{Effusio.} & \textbf{Infiltr.} & \textbf{Mass} & \textbf{Nodule} & \textbf{Pneunia.} & \textbf{Pneurax.}\\\midrule[0.2ex]
\begin{tabular}[c]{@{}c@{}}0.1 \end{tabular} &
\begin{tabular}[c]{@{}c@{}}\textbf{CAM}\\\textbf{LRP}\\\textbf{RAP} \end{tabular} &
\begin{tabular}[c]{@{}c@{}}    0.243\\\textbf{0.500}\\0.487\end{tabular} &
\begin{tabular}[c]{@{}c@{}}    0.336\\0.687\\\textbf{0.754}\end{tabular} &
\begin{tabular}[c]{@{}c@{}}    0.311\\\textbf{0.503}\\0.490\end{tabular} &
\begin{tabular}[c]{@{}c@{}}    0.309\\0.567\\\textbf{0.576}\end{tabular} &
\begin{tabular}[c]{@{}c@{}}    0.225\\\textbf{0.514}\\0.496\end{tabular} &
\begin{tabular}[c]{@{}c@{}}    0.182\\\textbf{0.435}\\0.416\end{tabular} &
\begin{tabular}[c]{@{}c@{}}    0.295\\0.568\\\textbf{0.572}\end{tabular} &
\begin{tabular}[c]{@{}c@{}}    0.259\\\textbf{0.456}\\0.447\end{tabular}
\\\midrule[0.2ex]
\begin{tabular}[c]{@{}c@{}}0.2 \end{tabular} &
\begin{tabular}[c]{@{}c@{}}\textbf{CAM}\\\textbf{LRP}\\\textbf{RAP} \end{tabular} &
\begin{tabular}[c]{@{}c@{}}    0.304\\\textbf{0.560}\\0.534\end{tabular} &
\begin{tabular}[c]{@{}c@{}}    0.439\\0.569\\\textbf{0.703}\end{tabular} &
\begin{tabular}[c]{@{}c@{}}    0.368\\\textbf{0.543}\\0.511\end{tabular} &
\begin{tabular}[c]{@{}c@{}}    0.369\\0.603\\\textbf{0.609}\end{tabular} &
\begin{tabular}[c]{@{}c@{}}    0.280\\\textbf{0.582}\\0.540\end{tabular} &
\begin{tabular}[c]{@{}c@{}}    0.238\\\textbf{0.506}\\0.461\end{tabular} &
\begin{tabular}[c]{@{}c@{}}    0.355\\0.602\\\textbf{0.606}\end{tabular} &
\begin{tabular}[c]{@{}c@{}}    0.303\\\textbf{0.494}\\0.471\end{tabular}
\\\midrule[0.2ex]
\begin{tabular}[c]{@{}c@{}}0.3 \end{tabular} &
\begin{tabular}[c]{@{}c@{}}\textbf{CAM}\\\textbf{LRP}\\\textbf{RAP} \end{tabular} &
\begin{tabular}[c]{@{}c@{}}    0.353\\0.563\\\textbf{0.565}\end{tabular} &
\begin{tabular}[c]{@{}c@{}}    0.525\\0.428\\\textbf{0.622}\end{tabular} &
\begin{tabular}[c]{@{}c@{}}    0.408\\\textbf{0.526}\\0.518\end{tabular} &
\begin{tabular}[c]{@{}c@{}}    0.413\\0.558\\\textbf{0.608}\end{tabular} &
\begin{tabular}[c]{@{}c@{}}    0.326\\\textbf{0.583}\\0.571\end{tabular} &
\begin{tabular}[c]{@{}c@{}}    0.286\\\textbf{0.545}\\0.496\end{tabular} &
\begin{tabular}[c]{@{}c@{}}    0.403\\0.564\\\textbf{0.608}\end{tabular} &
\begin{tabular}[c]{@{}c@{}}    0.339\\\textbf{0.492}\\0.479\end{tabular}
\\\midrule[0.2ex]
\begin{tabular}[c]{@{}c@{}}0.4 \end{tabular} &
\begin{tabular}[c]{@{}c@{}}\textbf{CAM}\\\textbf{LRP}\\\textbf{RAP} \end{tabular} &
\begin{tabular}[c]{@{}c@{}}    0.396\\0.543\\\textbf{0.569}\end{tabular} &
\begin{tabular}[c]{@{}c@{}}    \textbf{0.591}\\0.441\\0.544\end{tabular} &
\begin{tabular}[c]{@{}c@{}}    0.442\\0.502\\\textbf{0.510}\end{tabular} &
\begin{tabular}[c]{@{}c@{}}    0.451\\0.508\\\textbf{0.570}\end{tabular} &
\begin{tabular}[c]{@{}c@{}}    0.370\\0.552\\\textbf{0.573}\end{tabular} &
\begin{tabular}[c]{@{}c@{}}    0.328\\\textbf{0.561}\\0.521\end{tabular} &
\begin{tabular}[c]{@{}c@{}}    0.445\\0.515\\\textbf{0.576}\end{tabular} &
\begin{tabular}[c]{@{}c@{}}    0.370\\\textbf{0.484}\\0.480\end{tabular}
\\\midrule[0.2ex]
\begin{tabular}[c]{@{}c@{}}0.5 \end{tabular} &
\begin{tabular}[c]{@{}c@{}}\textbf{CAM}\\\textbf{LRP}\\\textbf{RAP} \end{tabular} &
\begin{tabular}[c]{@{}c@{}}    0.437\\0.519\\\textbf{0.548}\end{tabular} &
\begin{tabular}[c]{@{}c@{}}    \textbf{0.635}\\0.424\\0.480\end{tabular} &
\begin{tabular}[c]{@{}c@{}}    0.468\\0.485\\\textbf{0.493}\end{tabular} &
\begin{tabular}[c]{@{}c@{}}    0.483\\0.477\\\textbf{0.519}\end{tabular} &
\begin{tabular}[c]{@{}c@{}}    0.411\\0.520\\\textbf{0.547}\end{tabular} &
\begin{tabular}[c]{@{}c@{}}    0.369\\\textbf{0.551}\\0.527\end{tabular} &
\begin{tabular}[c]{@{}c@{}}    0.479\\0.482\\\textbf{0.525}\end{tabular} &
\begin{tabular}[c]{@{}c@{}}    0.397\\\textbf{0.479}\\0.478\end{tabular}
\\
\bottomrule[0.3ex]
\end{tabular}

\label{tab:2}
}
\end{table*}

\subsection{Results}
\subsubsection{Quantitative Assessment}
To validate the efficiency of the attribution methods in visualizing target lesions, we compared with CAM, which is widely used in medical fields to guarantee reliability. The heatmaps from each method are normalized in $\{0\sim 1\}$ and the threshold $T$ is applied. Negative attributions are cast as zero for the fair comparison. Tab.\ref{tab:2} shows the results of mean IOU per each class on CheXNet. Pixels that have lower relevance value than the threshold are cast as zero. As shown in Tab. \ref{tab:2}, CAM shows low IOU performance rather than LRP and RAP in the low threshold. Since the heatmaps from CAM are generated by resizing from low dimension feature maps to original input size, it is hard to visualize the delicate interpretations for the target lesions. As threshold value is increased, low attributions widely spread with irrelevant parts are deleted, resulting in improvement of the localization performance of CAM. On the contrary, the attribution methods: LRP and RAP show the decrement of IOU when the threshold is too high. After the output predictions are fully decomposed and mapped into pixel-by-pixel, attributions compose detail visual explanations with the degree of importance.

\subsubsection{Qualitative Assessment}
\begin{figure}[!t]
\includegraphics[width=\textwidth]{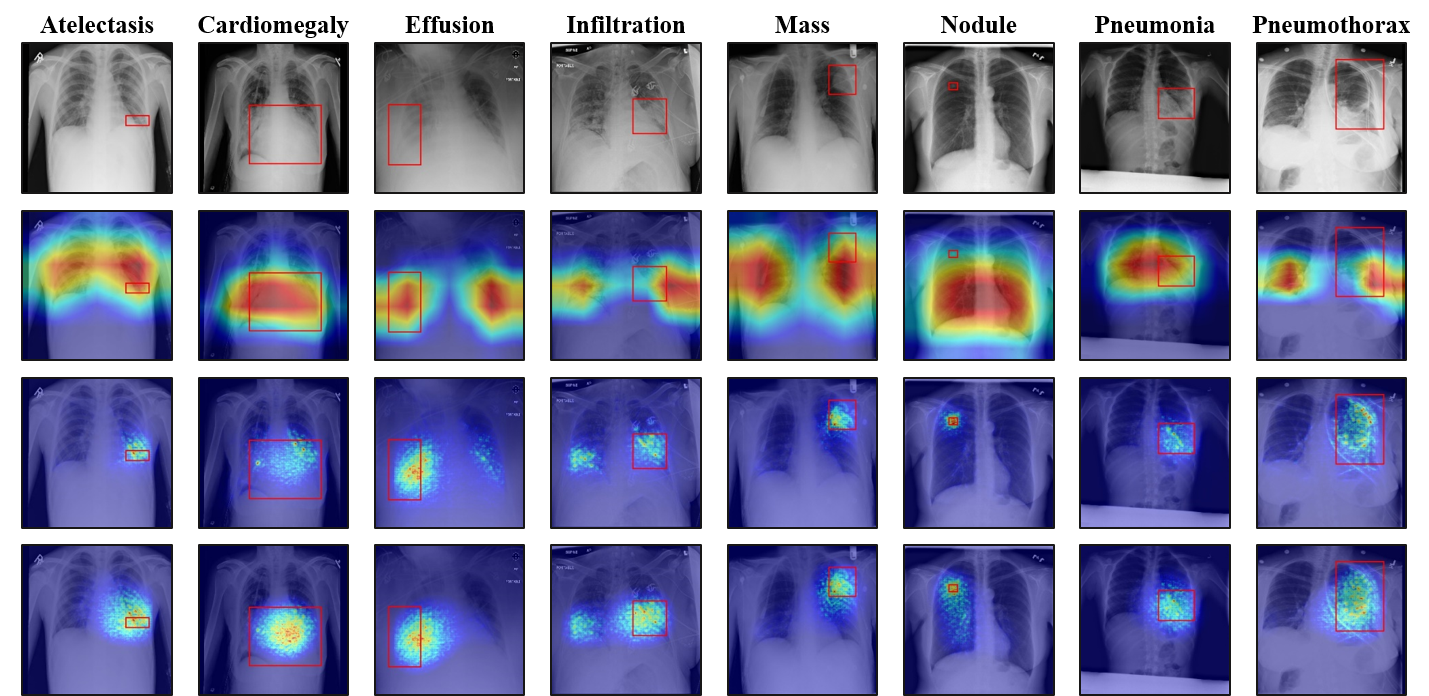}               
\caption{The qualitative comparison of visual explanations. Each row illustrates an X-ray image, CAM, LRP, and RAP, respectively. Red square indicates the bounding box for each disease.} \label{fig3}
\end{figure}

For qualitatively evaluating the heatmaps from each method, we compare the results by examining how the high activated points are distributed in the bounding box. As the methods have the same purpose for emphasizing the most important factors, we can assess whether each method is consistent in attributing positive relevance. Fig.\ref{fig3} presents the heatmaps from each method: CAM, LRP, and RAP for the diagnostic decisions by CheXNet. We qualitatively assessed all images in the test set of NIH chest X-ray 14 dataset, and most of them appear to show similarly satisfactory results in a human view. More qualitative comparisons are illustrated in the supplementary material.

\section{Inconsistency of Intention}
It is not trivial to elucidate the decision of DNNs because of the opacity from the myriad of linear and nonlinear operations. Tempted by the impressive performance, it is easy to believe that the criterion of decisions is from the same intentions as human. \cite{lapuschkin-ncomm19} points out this problem and insists the necessity of explaining techniques and their evaluation metrics. Especially in the medical field, the identification of causes for diagnosis is crucial to ensure reliability.

As described in Section 3.2, we trained DNN models on RSNA Pneumonia detection Datasets for binary classification of Pneumonia images. The performance of each model based on general learning methods shows fair performance. Fig.\ref{fig4} illustrates the visual explanations of what DNNs mainly focused on. The input X-ray images are correctly classified as target labels. For the Pneumonia x-ray, the relevance from trained models: VGG, Resnet, Densenet is distributed on irrelevant area of lesions (bounding box) without regular patterns. However, CheXNet, pretrained in NIH dataset with certain purpose to classify various diseases, shows clear visual explanations corresponding to Pneumonia. For the normal X-ray image, relevance from trained model appears in areas that support the normal lung's clear shape. The result of additional normal images also show this similar relevance patterns. Here, the Interesting phenomenon is that DNN models learn the status of normal lung images, not learn the characteristics of Pneumonia disease. Since we do not provide any supervisions of the lesion area, DNN focuses on the lungs in a normal state, which is rather large in volume and clearly visible to pursue higher performance. CheXNet classified this input X-ray as Cardiomegaly, which is not closely related to lung diseases, and the visual explanation clearly support the diagnostic decision by emphasizing the lesion area of heart.

\begin{figure}[!t]
\includegraphics[width=\textwidth]{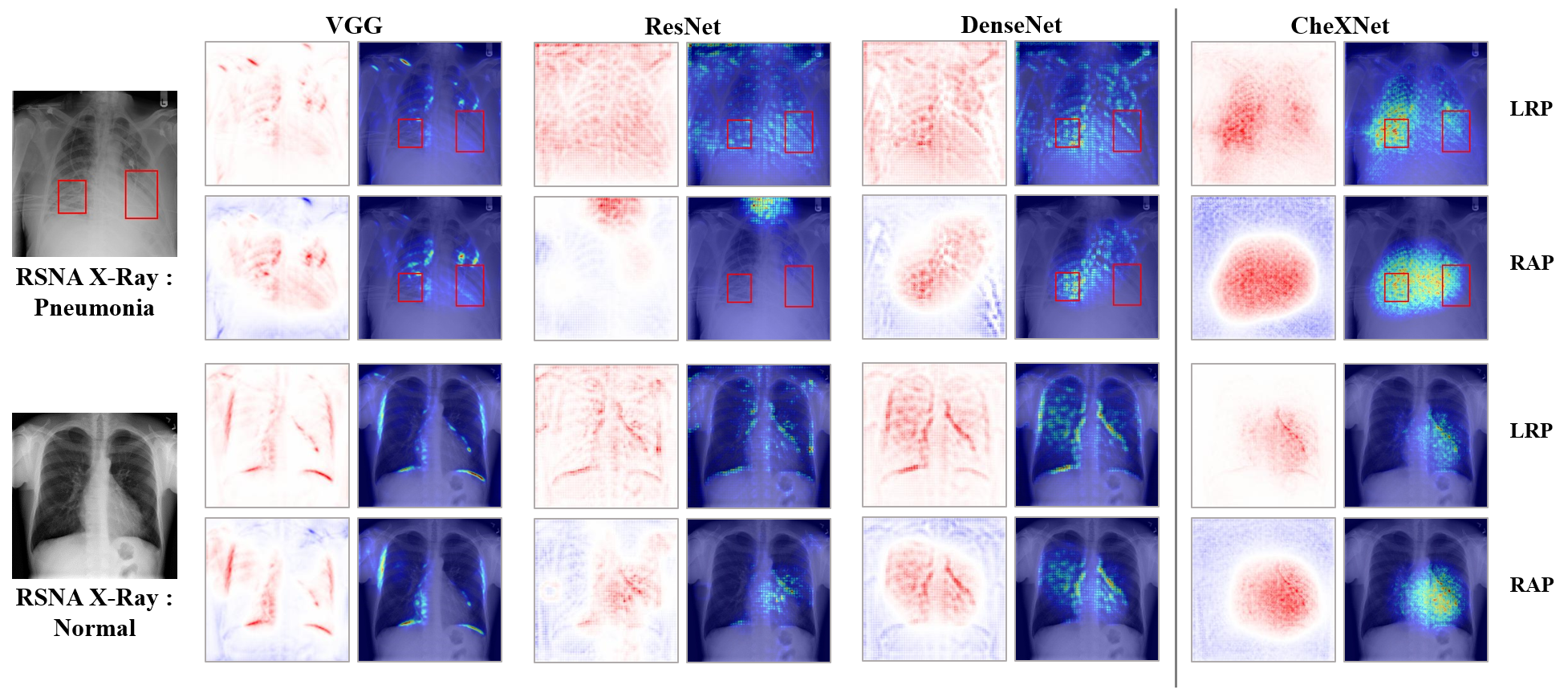}
\caption{Investigating the inconsistency between human intention and what DNN has learned. Please see Section 4 for details.} \label{fig4}
\end{figure}
\section{Conclusion}
In this paper, we demonstrate an efficient method to unmask the opacity of DNNs and provide interpretation of diagnostic decision by utilizing explaining techniques. The introduced methods: LRP and RAP can visualize more accurate and clear parts of lesions than generally used CAM. Generated heatmaps indicate the important factors for deciding the target diseases with intensity from high relevant to low relevant. We utilize chest X-ray datasets: NIH ChestX-ray14 and RSNA pneumonia datasets to verify how attribution methods could localize the target lesions without any supervision of bounding box. For the quantitative evaluation, we use mean Intersection of Union for the visualization methods: CAM, LRP and RAP. The results show that fully decomposing the network with investigating the contributions of neurons makes it possible to clearly localize the part of lesions. Furthermore, we analyze the inconsistency of human intentions and DNNs by utilizing explaining methods, and emphasize the necessity of interpretability for the adoption of machine intelligence in medical domain.
\begin{figure}[!t]
\includegraphics[width=\textwidth]{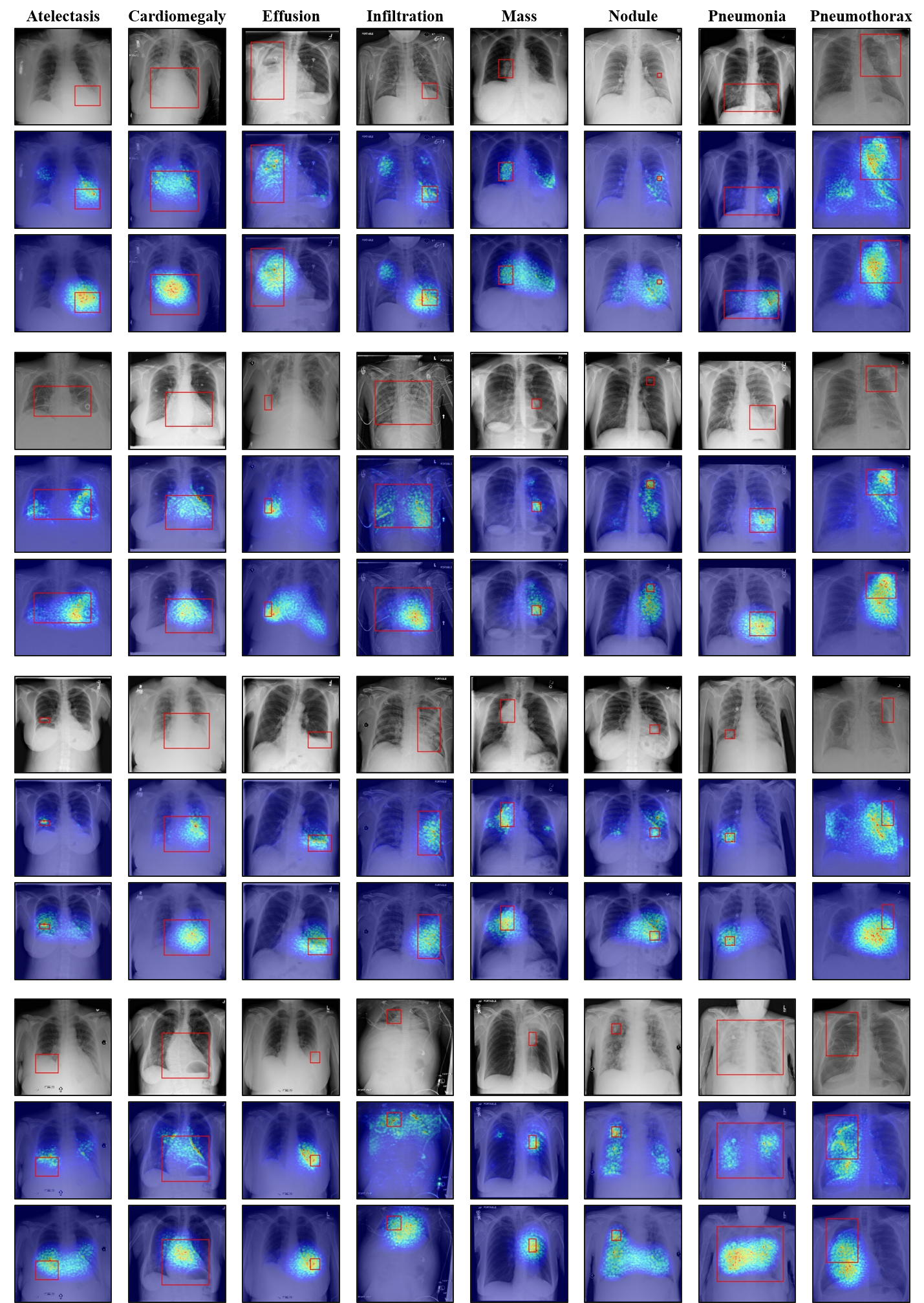}
\caption{Additional comparison of visual explanations generated from CheXNet. First, second, and third row in each tuple denote image, LRP, and RAP, respectively. The visualization is the result without applying threshold.} \label{fig4}
\end{figure}
\begin{figure}[!t]
\includegraphics[width=\textwidth]{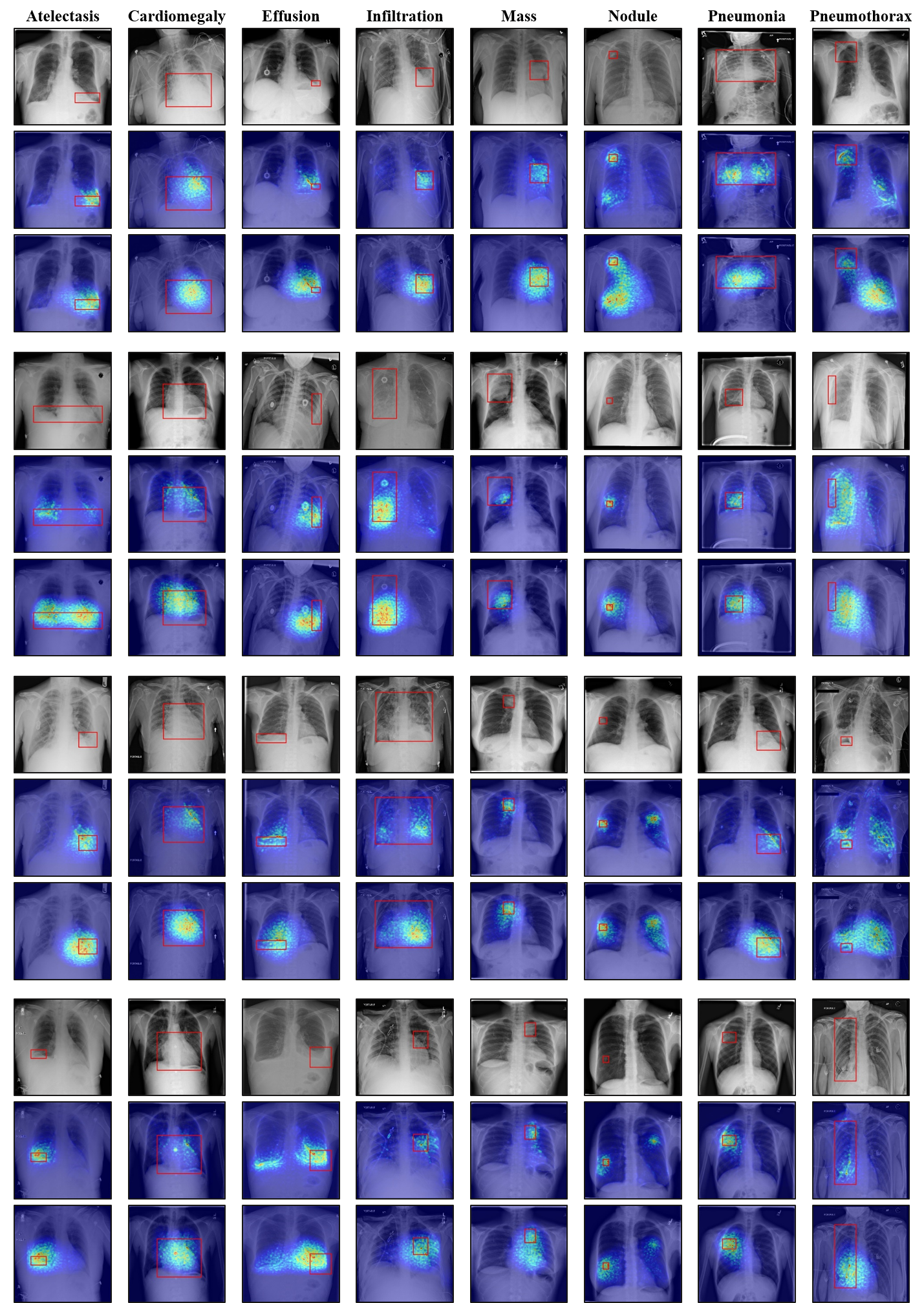}
\caption{Additional comparison of visual explanations generated from CheXNet. First, second, and third row in each tuple denote image, LRP, and RAP, respectively. The visualization is the result without applying threshold.} \label{fig4}
\end{figure}
%
%
\bibliographystyle{splncs04}
\bibliography{egbib}

\end{document}